\documentclass[apj,twocolumn]{openjournal}

\usepackage{graphicx}
\usepackage{float}
\usepackage[margin=10pt,font=small,labelformat=simple,labelsep=colon]{caption}
\usepackage{subcaption}
\usepackage{xcolor}

\usepackage{amsmath}
\usepackage{amsfonts}
\usepackage{amssymb}
\usepackage{upgreek}
\usepackage[pdfborder={0 0 0},colorlinks=false,linkcolor=black,citecolor=black]{hyperref}
\usepackage{txfonts}
\usepackage[utf8]{inputenc}
\usepackage{multirow}
\usepackage{booktabs}
\usepackage{cleveref}

\newcommand{\orcidauthor}[3]{\author{\href{http://orcid.org/#1}{#2 \openin1 Orcid-ID.png \ifeof1 \else \hskip2pt\includegraphics[width=9pt]{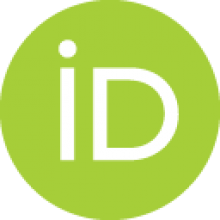}\fi}$^{#3}$}}
\newcommand{\citeg}[1]{\citep[e.g.][]{#1}}

\graphicspath{ {./figs/} }
\DeclareUnicodeCharacter{2212}{-}

\begin{document}

\title[A \textmu Hz PSR timing telescope?]{Investigating the viability of dedicated microhertz gravitational wave pulsar timing}

\orcidauthor{0009-0000-5199-3100}{C.~F.~Sheridan}{1\star}
\orcidauthor{0000-0002-4553-655X}{E.~F.~Keane}{1}

\affiliation{$^1$ School of Physics, Trinity College Dublin, College Green, Dublin 2, D02 PN40, Ireland}

\thanks{$^\star$Email: \href{mailto:csherid2@tcd.ie}{csherid2@tcd.ie}}

\begin{abstract}
The microhertz gravitational wave (GW) frequency regime remains one of the least explored regions of the GW spectrum. This is due to the physical limitations of our current and upcoming GW observatories such as LIGO and LISA, which are tuned to dHz and mHz respectively. Proposed experiments targeting the \textmu Hz regime are ambitious, expensive and unlikely to be operational within the next 20 years. Pulsar Timing Arrays (PTAs), with their Galaxy-sized baselines, occupy the other end of the spectrum; the frequencies they are sensitive to are a function of the observing cadence and total dataset timespan. Existing PTA campaigns have cadences and timespans corresponding to the nanohertz regime, however the possibility exists that they can be `pushed' to detect higher frequencies. This paper investigates the viability of a custom microhertz-sensitive PTA, observing a small number of millisecond pulsars with a daily cadence, over the course of 1 and 2 years respectively. We assume the use of both 100-m and 500-m class radio telescopes for these observations, and an idealised noise model with negligible systematics to determine the most optimistic detection threshold. We compute the sensitivity of our proposed high-cadence PTA configurations to power-law signals and individual GW sources. We find that a 500-m class telescope would likely be able to set new upper limits on individual sources and GW background signals in the $0.1-1$~\textmu Hz range; while a 100-m class telescope would, in the best case, only match that of the performance of the EPTA second data release in the same range. Finally we consider a possible \textmu Hz PTA consisting of SKA-Mid and DSA observing 24 pulsars; we find this performs on par with the 500-m class while using far less telescope time, presenting a more feasible solution to microhertz-GW pulsar timing.
\end{abstract}

\maketitle


\section{Introduction}

Pulsars are rapidly rotating neutron stars, that generally emit radiation across the radio spectrum, and appear, observationally, as pulsing objects, one pulse per rotation, as seen by radio telescopes on Earth \citep{Shapiro_Teukolsky}. The fastest-spinning sub-group of these objects are the millisecond pulsars (MSPs); these have been `spun up' to rotational periods of the order of milliseconds through angular momentum transfer interactions with binary partners~\citeg{Backer_1982, Bhattacharya_1991}. MSPs have extremely stable pulse profiles and timing properties~\citep{Liu_2012, Ghosh_2025} and therefore can be used as highly accurate clocks, rivalling the precision of some of the best atomic clocks on Earth~\citep{Hobbs_2020}. Their exceptional stability allows for precise timing measurements. Observations with an ensemble of pulsars can be used as a gravitational wave (GW) detector \citep{Sazhin_1978,Detweiler_1979}, with the Earth and pulsar forming either end of a detection arm sensitive to very long-wavelength GWs. Timing is done by timestamping detected pulses, forming a collection of Times of Arrival (ToAs) for a given pulsar; observed ToAs are then subtracted from the corresponding predicted ToAs using a pulsar timing model, forming the so-called `residuals' \citep{Detweiler_1979}. After accounting for the star's rotational evolution, binary features, propagation effects etc., remaining non-zero residuals can be inferred as having been caused by passing GWs between Earth and the pulsar. Multiple MSPs can be used to form a combined detector with many Galaxy-sized arms, known as a Pulsar Timing Array \citep[PTA;][]{Foster_1990,Hellings_1983}.

The minimum and maximum GW frequency of a PTA can be calculated from two parameters, namely the total observation time $T_\mathrm{obs}$ and the total number of ToAs $N$ \citep{Jenet_2005,Hobbs_2009}, as:
\begin{align}
f_\text{min} &\approx \frac{1}{T_\text{obs}}\;, \label{eq:fmin}\\
f_\text{max} &\approx f_\text{Nyq} = \frac{N}{2T_\text{obs}}\;, \label{eq:fmax}
\end{align}
where $f_{\rm Nyq}$ is the Nyquist frequency. This however assumes uniform ToA sampling; PTAs unequally sample ToAs, given real-world telescope operational constraints, which in turn contributes to a higher upper frequency limit \citep{Koen_2006}. Notwithstanding, we can use \eqref{eq:fmin} and \eqref{eq:fmax} as reasonable estimates of the frequency range wherein a pulsar timing experiment is sensitive to GWs. PTAs generally have data sets spanning decades, observing with a weekly to monthly cadence~\citep{EPTA1_2023,nanoGrav_2023,PPTA_2023}, so correspondingly they are sensitive to GWs with frequencies in the range of $10^{-9}-10^{-7}$~Hz. Meanwhile, Earth-based detectors, such as LIGO, operate in the range $10-10^3$~Hz \citep{aLIGO_2015}. By increasing the cadence of PTA observations to a daily rate, one can push the upper frequency limit into \textmu Hz frequencies. 

\begin{figure*}
    \centering
    \includegraphics[width=0.89\linewidth]{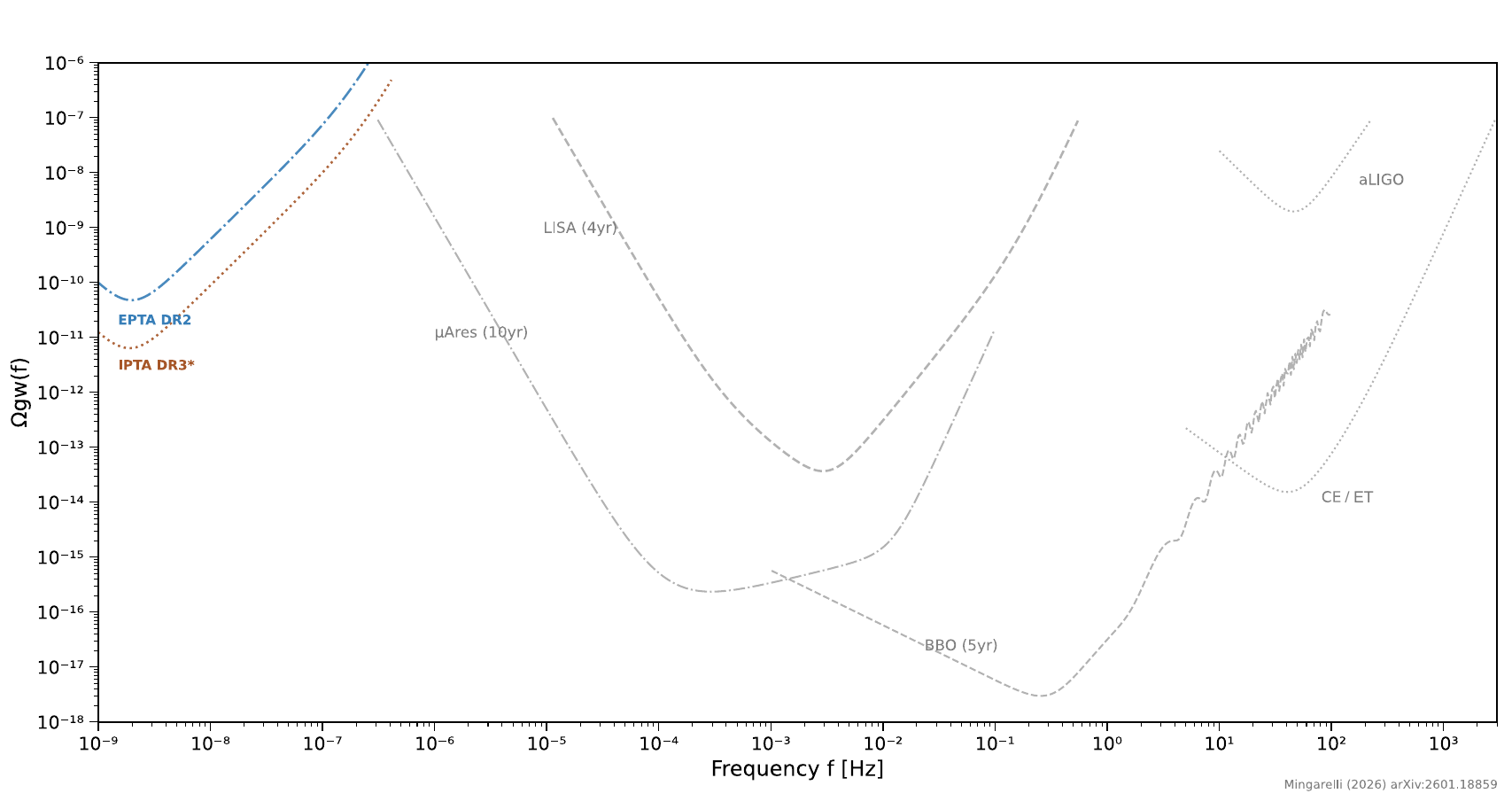}
    \caption{The sensitivity curves of various current (EPTA DR2, aLIGO), upcoming (LISA, IPTA DR3, Einstein Telescope/ET) and proposed (Cosmic Explorer/CE, Big Bang Observer/BBO, \textmu Ares) GW detectors to a stochastic GWB in units of energy density. Note the current gap in the frequency coverage at the \textmu Hz regime. Figure generated from an online tool created by \citet{Mingarelli_2026a}, see \href{https://gwbceilings.streamlit.app/}{\color{cyan}{here}}.}
    \label{fig:lit_gw_scurves}
\end{figure*}

The motivation for doing this is two-fold. Firstly, current PTA sensitivity is optimised to a frequency range where there is a degeneracy between possible contributions to a Gravitational Wave Background (GWB): from improperly-modelled Super-Massive Black Hole Binaries (SMBHBs) and from cosmological sources (also called `New Physics') \citep{nanoGrav_2023, EPTA4_2024}. While this degeneracy will likely be broken with further improvements to the existing PTA collaborations and methods \citeg{Goncharov_2026}, a potential concurrent solution would be to probe the microhertz GW frequency range, where SMBHBs are predicted not to radiate with the same strength as in the nanohertz range. There also exists microhertz-specific signals of interest, such as the gravitational wave signature of the QCD phase transition \citep{Renzini_2022}, which itself is a strong motivator to explore this regime. Secondly, there has been very limited exploration of the \textmu Hz regime, owing to limitations in our detectors (see Fig. \ref{fig:lit_gw_scurves}), meaning it is potentially an as yet untapped discovery space. The Laser Interferometer Space Array (LISA) will eventually begin to fill in the gap from the millihertz side; with a planned launch date of 2035~\citep{LISA_2017}. Similarly, there are proposed space-based interferometers targetting the microhertz range specifically, such as \textmu Ares, with planned operation in the 2040s~\citep{muAres_2021}, and another which proposes the use of asteroids as GW test masses \citep{Fedderke_2022}. Proposed next generation Earth-based GW observatories include the Einstein Telescope \citep{ET_2010} and Comsic Explorer \citep{CE_2019} and target frequencies of Hz to kHz, with operation dates after 2035. There is an opportunity to attempt pulsar timing experiments in unexplored GW regimes between now and when these observatories come online, and to also provide some early insight into the microhertz regime, which could inform the  designs of these purpose-built telescopes, each scientific megastructures in their own right.

This paper explores the performance of a purpose-designed high-cadence PTA experiment exploring the \textmu Hz GW frequency regime, compared to its existing counterparts, and if it would be theoretically capable of detecting and/or constraining various cosmological GW background signals. We also explore its ability to detect individual sources of GWs. It is organised as follows: in \S\ref{sec:bckgrnd} we recap the potential sources of GWs that we expect in this frequency regime and outline the hypothetical observation campaign; in \S\ref{sec:mthdlgy} we discuss the formalism used to compute sensitivity curves and pulsar noise; \S\ref{sec:rslts-discssn} presents and discusses the power-law sensitivity curves and sky-map sensitivity of a hypothetical \textmu Hz-tuned PTA. For the remainder of this paper, natural units will be used, i.e. $k_{\rm B}=c=\hbar=1$, such that temperature and mass are expressed in terms of energy units. We take the Hubble constant to be $H_0=67.66$~km~s$^{-1}$~Mpc$^{-1}$, from Planck 2018 \citep{Plank2018}, also given in the dimensionless form $h_{100}=H_0/100~\text{km}~\text{s}^{-1}~\text{Mpc}^{-1}=0.6766$.

\section{Background} \label{sec:bckgrnd}
\subsection{GW Sources}
A gravitational wave background is predicted to be generated from the population of supermassive black hole binaries, peaking in the nHz frequency regime. The population of SMBHBs emitting \textmu Hz GWs is expected to be drastically less than those emitting in the nHz regime, due to the fact that the time a SMBHB spends in a GW frequency bin, $\tau_{\rm res}$, is proportional to $f_{\rm gw}^{11/3}$ \citep{Peters_1964}. This in turn means no stochastic background is expected from SMBHBs in the \textmu Hz region \citep{Afzal_2023,Sesana_2025}, as only a handful of SMBHBs would populate a given frequency bin in this regime and would be measurably transient on PTA timescales \citep{Mingarelli_2026b}. Without a stochastic GW background due to SMBHB, any detection could be confidently attributed to alternative cosmological sources, demonstrating their existence and providing valuable insight into how the early Universe may have formed. The lack of a stochastic background of SMBHB binaries could also make detecting individual bright binaries more tractable.  

\subsubsection{First Order Phase Transitions}\label{sec:FOPT}
First order phase transitions (FOPTs) are predicted to have occurred during the early evolution of our Universe, as various fields decoupled from one-another \citep{Witten_1984,Hogan_1986}. `First-order' in this case indicates that transitions would have begun at random isolated points and nucleated at relativistic speeds outwards as spherical bubbles. While the growing spherical bubbles would not have generated GWs, the interaction of bubbles with each other and the turbulence generated in the surrounding plasma would have produced a GW signature. The frequency of this signature is proportional to the energy scale of the transition $T_*$ \citep{Caprini_2008}, and the relation is given as
\begin{align} \label{eq:fotp_f_peak}
    f^\text{FOPT}_{\rm peak}\sim10^{-6}~\text{Hz}\cdot\frac{T_*}{200~\text{MeV}}
\end{align}
The QCD phase transition is the predicted moment where the quark-gluon plasma cooled sufficiently to coalesce into hadrons. The temperature of this transition is calculated to be on the order of $200~\text{MeV}$ \citep{Petreczky_2012,Kapusta_2003}. This corresponds to a time on the order of 10~\textmu s after the Big Bang in standard cosmology, which places it after the Electroweak transition but before Big Bang nucleosynthesis \citep{Kolb-Turner_1990}. The peak frequency of the QCD transition is approximately $1\upmu$Hz according to \eqref{eq:fotp_f_peak}, making it a target of a \textmu Hz search.

\subsubsection{Cosmic Strings}
Related to FOPTs are cosmic strings, which are 1-dimensional topological defects in spacetime that should arise at the interface between phase transition bubbles \citep{Nielsen_1973,Kibble_1976}. They are emitters of GWs, and it is predicted that a network of cosmic strings would contribute to the GWB \citep{Damour_2001,Blanco-Pillado_2017}. The main parameter determining the strength of a cosmic string background is the dimensionless cosmic string tension $G\mu$. \cite{Leclere_2024} constrains this to $G\mu \lesssim 10^{-9.9}$ for a particular cosmic string model, using a six-pulsar subset of the European PTA (EPTA) Data Release 2 \citep{EPTA1_2023}. The correct model for a cosmic string GWB remains an open question, and it is uncertain if a power law is sufficient to capture its signature, or whether something like a broken power law would be more accurate \citep{Caprini_2019}. We assume our desired target is a power law for reasons of simplicity, taking advantage of the boost in sensitivity this yields, explained in \S\ref{sec:scurves}.

\subsubsection{Domain Walls}
Domain Walls are another topological defect that may have been produced by the aforementioned phase transitions in the early Universe and are a two-dimensional analogue to Cosmic Strings \citep{Kibble_1976}. If they indeed formed they must have become unstable and annihilated themselves once the Universe cooled to a temperature $T_\text{ann}$ \citep{Zeldovich_1974}, and it is this self-annihilation that would produce a detectable GW signature, dependant on the value of $T_\text{ann}$. The other parameter determining their emission strength is their surface tension $\sigma_\text{DW}$ (see \citealt{Saikawa_2017} for a review of domain walls). We follow the description of domain walls given in \cite{Saikawa_2017}, which give the peak frequency and energy density of their emission spectrum to be
\begin{align}
    f^\text{DW}_\text{peak}&\approx1.1\times10^{-7}~\text{Hz}\cdot\left(\frac{g_{\rm *e}}{10}\right)^{1/2}\left(\frac{g_{\rm *s}}{10}\right)^{-1/3}\left(\frac{T_\text{ann}}{1~\text{GeV}}\right) \label{eq:dw_f_peak}\\
    \Omega_{\rm gw}^{\rm peak}&\approx\frac{10^{-26}}{h_{100}^2}\cdot\left(\frac{g_{\rm *e}}{10}\right)^{-4/3}\left(\frac{\sigma_{\rm DW}}{\rm 1~TeV^3}\right)^2\left(\frac{T_{\rm ann}}{1~\text{GeV}}\right)^{-4}\label{eq:dw_omega_peak}
\end{align}
Where $g_{\rm *e}$ and $g_{\rm *s}$ are  the relativistic energy and entropy degrees of freedom, both functions of temperature, evaluated at $T_\text{ann}$, and $h_{100}$ is the dimensionless Hubble constant. We consider an annihilation temperature of $T_\text{ann}=1~\text{GeV}$, corresponding to a time of 1 \textmu s after the Big Bang \citep{Kolb-Turner_1990}. We choose this as it corresponds to $g_{\rm *e}\approx10\approx g_{\rm *s}$ \citep{Husdal_2016}, meaning $f_\text{peak}^\text{DW}\sim10^{-7}~\text{Hz}$, thus making it a target for $\upmu$Hz GW pulsar timing.

\subsubsection{Individual Sources}
In addition to all-sky sensitivity, PTAs can also be `pointed' in particular directions on the sky, so as to detect GWs from monochromatic individual sources \citeg{EPTA5_2024}; such as continuous GW emission from bright individual SMBHBs, or in the context of cosmological sources, GW bursts generated by cosmic strings \citep{Damour_2000}. We find the possibility of detecting individual SMBHB, also known as Continuous Gravitational Waves (CWs) a particularly interesting avenue of investigation in this regime, as there have already been searches of high-cadence subsets of PTA datasets for such targets, using both WSRT and FAST \citeg{Yi_2014,Perera_2018,Caballero_2025,Zou_2026}. We use the upper bound on the strain amplitude of optimal-location SMBHB, given by \cite{Zou_2026}, as a benchmark for the required sensitivity needed to begin probing individual GW sources in the microhertz regime. They give this as $8.9\times10^{-14}$ at $1~$\textmu Hz, which was established using a high-cadence subset of observations of PSR~J1909$-$3744 from the second IPTA Data Release.

\begin{table}[]
    \caption{The parameters and corresponding values used in equation \eqref{eq:sigma_rms} for a WSRT-  and FAST-like observatory. The effective area of WSRT is given for fourteen 25-m dishes with a 40\% aperture efficiency. We take the FAST values from \cite{Jiang_2019} and \cite{Xu_2025}. The effective area of FAST is given for a 300-m illuminated aperture with 50\% efficiency. Any row marked with an asterix indicates the value was chosen by the authors.}
    \centering
    \footnotesize
    \begin{tabular}{c|c|c|c|c}
        \toprule
        \multicolumn{5}{c}{\textbf{Telescope Parameters}} \\
        \midrule
        Symbol & Name & WSRT Val. & FAST Val. &\\
        \midrule
        $f_\text{obs}$ & Observation Frequency & $340~$MHz & 1250~MHz & *\\
        $A_\text{eff}$ & Effective Area & $2750~\text{m}^2$& $35340~\text{m}^2$\\
        $t_\text{int}$ & ToA Integration Time & $3600~$s & 3600 s & *\\
        $n_\text{pol}$ & \# of Polarisations & 2 & 2 & \\
        $\Delta f$ & Telescope Bandwidth & $70~$MHz & 500 MHz & \\
        $T_\text{sys}$ & System Temperature & 250 K & 50 K & *\\
        \bottomrule
    \end{tabular}
    \label{tab:wsrt_props}
\end{table}

\subsection{Observation Campaign}
We assume a Westerbork Synthesis Radio Telescope (WSRT)-like observatory for the sensitivity calculations undertaken in this paper, using the 92-cm receiver and all fourteen 25-m dishes. WSRT serves as a good representation for a 100-metre class radio telescope; furthermore it is potentially on the cusp of an upgrade so that investigating the \textmu Hz PTA question is also a feasibility study for a potential science driver for such an upgrade. We also repeated our analyses using Five-hundred-metre Aperture Spherical Telescope (FAST)-like parameters, which has approximately 1 order of magnitude better gain \citep{FAST_2011}. The parameters we use for each of these telescopes are included in Table \ref{tab:wsrt_props}. For the simulated PTA, we choose four `good timers'\footnote{Pulsars with the smallest calculated radiometer error, $\sigma_\text{rad}$, for the observational setup.} from the EPTA DR2\footnote{\url{https://www.epta.eu.org/epta-dr2.html}} list, see Table \ref{tab:psr_list}. 

We present three observational campaigns: Campaign A, with a cadence of 2 ToAs per pulsar per day, assuming an hour-long observation to be integrated to form each ToA, corresponding to a daily 8 hours of observation, and a total observation baseline length of 1 year. Campaign B has a cadence of 4 ToAs per pulsar per day, with the same integration time, equating to a daily 16 hours of observation, with a total baseline of 2 years. All ToAs are assumed to be sampled at equal intervals. Campaigns A and B correspond approximately to a Nyquist upper frequency limit of $f_\text{max}\approx10~\upmu{\rm Hz,}$ and $20~\upmu$Hz respectively. Both these campaigns, especially B, are extremely intensive (Campaign B requires 11,680 hours on WSRT/FAST), so we investigate a third configuration, Campaign C, involving the upcoming Square Kilometer Array \cite[SKA;][]{Shannon_2025}, specifically SKA-Mid, and the Deep Synoptic Array \cite[DSA;][]{Hallinan_2019}. These are both next-generation interferometric radio arrays with substantial time dedicated towards pulsar timing. We assume the DSA will observe 12 MSPs each at a 1 day cadence, and choose a radiometer noise $\sigma^{\rm DSA}_{\rm rad}=50~{\rm ns}$, as an order-of-magintude representation of the DSA\footnote{Using the DSA Sensitivity Calculator (\url{https://www.deepsynoptic.org/sensitivity/index.html}).} \citep{Hallinan_2019}. For simplicity we assume the SKA will also observe 12 MSPs with the same cadence; and we take $\sigma^{\rm SKA}_{\rm rad}=50~{\rm ns}$. We assume this requires an integration time of 30 minutes for each ToA for both the DSA and SKA-Mid. In total this adds up to be 6 hours per day per telescope. We believe both these telescopes will be capable of simultaneous timing of multiple sources, and so consider the actual total observation time for this campaign to be less; assuming both of them use 4 dedicated tied-array beams simultaneously, this would correspond to a total 1.5 hour per day per telescope. We assume a total observation length of 2 years. We do not select 24 MSPs but rather generate random sky-positions for each as these same telescopes will themselves identify a substantial population of currently unknown MSPs. We believe this setup to be a feasible timing regime for these two observatories, and works hand-in-hand with their proposed timing projects, as these ToAs can be included in their nanohertz-targeting timing.

\begin{table}[]
    \caption{The pulsar ensemble used in this paper and corresponding properties $P$ (pulsar period), $W_{50}$ (pulse width at 50\% of the maximum) and $S_\text{mean}$ (the mean flux density at 400~MHz, 1400~MHz), taken from the ATNF catalogue. The quantity $\sigma^\text{WSRT}_\text{rad}$ is the radiometer error computed via Equation \eqref{eq:sigma_rms} for a WSRT-like telescope using $S_{\rm mean}^{400}$. We use the same pulsars for the FAST-like sensitivity curve, computing $\sigma^{\rm FAST}_\text{rad}$ with the modified telescope parameters in Table \ref{tab:wsrt_props} and $S_{\rm mean}^{1400}$.}
    \centering
    \footnotesize
    \begin{tabular}{c|c|c|c|c|c|c}
        \toprule
        \multicolumn{7}{c}{\textbf{Pulsar Array}} \\
        \midrule
        Pulsar & \shortstack{$\sigma^\text{WSRT}_\text{rad}$\\(ns)} & \shortstack{$\sigma^\text{FAST}_\text{rad}$\\(ns)} &\shortstack{$P$\\(ms)} & \shortstack{$W_{50}$\\(ms)} & \shortstack{$S^{400}_\text{mean}$\\(mJy)} & \shortstack{$S^{1400}_\text{mean}$\\(mJy)}\\
        \midrule
        J1012+5307 & 586 & 26 & 5.26 & 0.2 & 30 & 4\\
        J1022+1001 & 522 & 58 & 16.45 & 0.5 & 75 & 3.9\\
        J1744$-$1134 & 389 & 16 & 4.07 & 0.1 & 18 & 2.6\\
        J1857+0943 & 3543 & 83 & 5.36 & 0.5 & 20 & 5.0\\
        \bottomrule
    \end{tabular}
    \label{tab:psr_list}
\end{table}

\section{Methodology} \label{sec:mthdlgy}
\subsection{Sensitivity Curves}\label{sec:scurves}
In this section we follow the formalism derived by \cite{Thrane_2013} to characterise the sensitivity of a PTA to a power-law signal. We provide a brief summary of their derivation below. The expected power signal-to-noise ratio (SNR) $\rho$, for a cross-correlation search for an unpolarised and isotropic stochastic gravitational wave background is
\begin{align}\label{eq:pta_snr}
\rho = \sqrt{2T_{\text{obs}}}\left[ \int_{f_\text{min}}^{f_\text{max}}\sum_{\rm i=1}^{M}\sum_{\rm j>i}^M\frac{\Gamma_{\rm ij}^2(f)S_{\rm h}^2(f)}{P_{\rm ni}(f)P_{\rm nj}(f)}df\right]^{1/2}\;,
\end{align}
where $M$ is the number of GW detectors, in this case pulsars. For two given detectors $i$,~$j$, $\Gamma_{\rm ij}$ is the overlap reduction function between them and $P_{\rm ni}$ and $P_{\rm nj}$ are the respective noise power spectral densities. The gravitational wave power spectral density from a target background is $S_{\rm h}$. From this one can infer an effective strain noise power spectral density
\begin{align}
    S_\text{n}(f)\equiv\left[\sum_{i=1}^{M}\sum_{j>i}^M\frac{\Gamma_{\rm ij}^2(f)}{P_{\rm ni}(f)P_{\rm nj}(f)}\right]^{-1/2}\;,
\end{align}
and the corresponding noise characteristic strain $h_\text{n}$ and noise energy density $\Omega_\text{n}$ can be calculated using the following relations
\begin{align}
    \label{eq:h_c}
     h_{\rm n}(f)&=\sqrt{fS_{\rm n}(f)}\;, \\
    \label{eq:omega_gw}
    \Omega_{\rm n}(f)&=\frac{2\pi^2}{3H_0^2}f^3S_{\rm n}(f)\;, 
\end{align}
respectively\footnote{See \cite{Moore_2015} for explanation of the various GW units.}. When the signal of interest follows a power law, one can integrate over frequency to account for the higher sensitivity of a PTA to power-law signals
\begin{align} \label{eq:pl_energy-density}
    \Omega_\alpha=\frac{\rho}{\sqrt{2T_\text{obs}}} \left[ \int_{f_\text{min}}^{f_\text{max}} \frac{(f/f_\text{ref})^{2\alpha}}{\Omega_\text{n}^2(f)}df \right]^{-1/2}\;.
\end{align}
For a set of indices $\alpha$, the power-law integrated sensitivity curve is then given as
\begin{align}\label{eq:sensitivity_curve}
    \Omega_\text{gw}&=\max_\alpha\left[\Omega_\alpha\left(\frac{f}{f_\text{ref}}\right)^\alpha\right]\;.
\end{align}
This can be converted to other forms using \eqref{eq:h_c}, \eqref{eq:omega_gw}. For a PTA, the overlap reduction function is given as:
\begin{align}
    \Gamma_{\rm ij}(f)=\frac{1}{3(2\pi f)^2}\zeta_{\rm ij}\;,
\end{align}
where
\begin{align}
    \zeta_{\rm ij} \equiv&\frac{3}{2}\left(\frac{1-\cos\theta_{\rm ij}}{2}\right)\log\left(\frac{1-\cos\theta_{\rm ij}}{2}\right)\nonumber \\ 
    &-\frac{1}{4}\left(\frac{1-\cos\theta_{\rm ij}}{2}\right)+\frac{1}{2}+\frac{1}{2}\delta_{\rm ij}
\end{align}
is the Hellings-Downs factor \citep{Hellings_1983} such that it is normalised for a single pulsar ($\zeta_{\rm ii}=1$). When characterising PTA sensitivity to individual sources of GWs, we cannot use the formalism thus far developed, but refer the reader to \cite{Hazboun_2019b}, which details the method used. In short, we compute $S_{\rm n}$ for each sky position on the ecliptic and calculate the strain amplitude\footnote{Note that characteristic strain and strain amplitude are different GW measurements \citep{Moore_2015}.} $h_0(\hat{n})$ of a source at sky position $\hat{n}$ emitting GWs at a frequency of $f_0$ that would yield a signal-to-noise ratio of $\rho$, as
\begin{align} \label{eq:skymap_snr}
    h_0(\hat{n})=\rho\sqrt{\frac{S_{\rm n}(f_0,-\hat{n})}{T_{\rm obs}}}\;.
\end{align}

The library \texttt{hasasia} \citep{Hazboun_2019a} was developed to compute sensitivity curves and sky maps of a given PTA without the need to use computationally expensive simulations. It implements the formalism detailed above. We use this library to produce data in Figs. \ref{fig:computed_gw_scurves1}, \ref{fig:computed_gw_skymap1}, \ref{fig:dsa-ska}. We verified the power-law sensitivity output of \texttt{hasasia} by independently developing software to compute Equation \eqref{eq:sensitivity_curve}. After successfully verifying the calculations we proceeded using \texttt{hasasia} due to its additional features and polish.

\subsection{Pulsar Timing Noise}
Like all experimental apparatus, PTAs have a noise component that must be sufficiently modelled to achieve desired sensitivity. This noise can arise from the pulsar itself, the intermediate space between it and Earth, observatory/instrumentation/clock noise and errors arising from outdated ephemerides. Broadly speaking, PTA noise can be divided into two categories, time-uncorrelated ``white'' noise and time correlated ``red'' noise \citep{Lentati_2016}.

For the purposes of this feasibility study we are only concerned with order of magnitude predictions of sensitivity, we assume a very simplified model of white noise; following \cite{Thrane_2013}, we take the power spectrum to be 
\begin{align}
    P_\text{WN}=2\Delta t \sigma_\text{rad}^2\;,
\end{align}
where $\Delta t = 1/\text{cadence}$ and $\sigma_\text{rad}$ is the radiometer error. For each pulsar we calculate $\sigma_\text{rad}\approx W/\text{SNR}$ \citep{Backer_1989}, which we expand as
\begin{align} \label{eq:sigma_rms}
    \sigma_\text{rad}\approx\beta\frac{2k_BT_\text{sys}}{A_\text{eff}S_\text{mean}\sqrt{t_\text{int}n_\text{pol}\Delta f}}\sqrt{\frac{W^3}{p-W}}\;,
\end{align}
using the radiometer equation~\citep{PulsarHandbook}. The `correction factor' $\beta$ accounts for various effects such as digitisation loss, pulse shape mismatch, etc., which we take as $\beta=1.25$ as a catch-all. The remaining variables in \eqref{eq:sigma_rms} are given in Tables \ref{tab:wsrt_props} and \ref{tab:psr_list}, and the computed values for $\sigma_\text{rad}$ are given in Table \ref{tab:psr_list}. We use pulsar properties taken from version 2.7.0 of the ATNF pulsar catalogue\footnote{\url{https://www.atnf.csiro.au/research/pulsar/psrcat/}} \citep{ATNF_2005}.

Red noise is modelled as a power law \citep{Lentati_2016}, i.e.
\begin{align}
    P_\text{RN}\propto A_\text{RN}\left(f\right)^{-\gamma_\text{RN}}\;,
\end{align}
where $A_\text{RN}$ and $\gamma_\text{RN}$ are respectively the amplitude and spectral index of a given red noise process. Each red noise process will have a different value for each of these parameters, and one or more of these may couple with and muffle a GWB power-law signature. We omit red noise from our analysis and so we present best-case estimates. Depending on the pulsar, jitter is also a substantial factor to consider for a telescope as sensitive as FAST \citeg{Liu_2012,Wang_2024}, but we likewise omit it for simplicity.

\section{Results \& Discussion} \label{sec:rslts-discssn}
\subsection{Power-Law Sensitivity} \label{sec:rslts_subsec1}
We include GW curves from the three cosmological sources previously mentioned (phase transitions, cosmic strings and domain walls) as possible GW background targets in this regime. We use simulated data for a cosmic string-generated stochastic GWB from \cite{Blanco-Pillado_2017}. This simulation modelled the growth and scaling of a cosmic string network using the Nambu-Goto approximation, to compute its expected GW profile. We select a string tension of $10^{-10}$, rounding the $10^{-9.9}$ upper limit given in \cite{Leclere_2024}. We also include the predicted GW signature of a QCD phase transition (corresponding to $T_\text{ann}\approx200~{\rm MeV}$) taken from \cite{Renzini_2022}. Finally we compute a domain wall background signal using results of numerical simulations \citeg{Hiramatsu_2010,Hiramatsu_2014,Kawasaki_2011} which show that 
\begin{align*}
    \Omega_{\rm gw}\propto\left\{
    \begin{array}{ll}
         f^3~~{\rm for}~f<f_{\rm peak}  \\
         f^{-1}~~{\rm for}~f>f_{\rm peak} 
    \end{array},
    \right.
\end{align*} 
with $f_{\rm peak}$ given by \eqref{eq:dw_f_peak}. We assume $\sigma_{\rm DW}^{1/3}=10^6~$GeV and $T_{\rm ann}=1~$GeV, which are selected from the allowable region of the parameter space given in \cite{Saikawa_2017}, and the peak energy density is then calculated using \eqref{eq:dw_omega_peak}.

\begin{figure}
    \centering
    \includegraphics[width=0.98\linewidth]{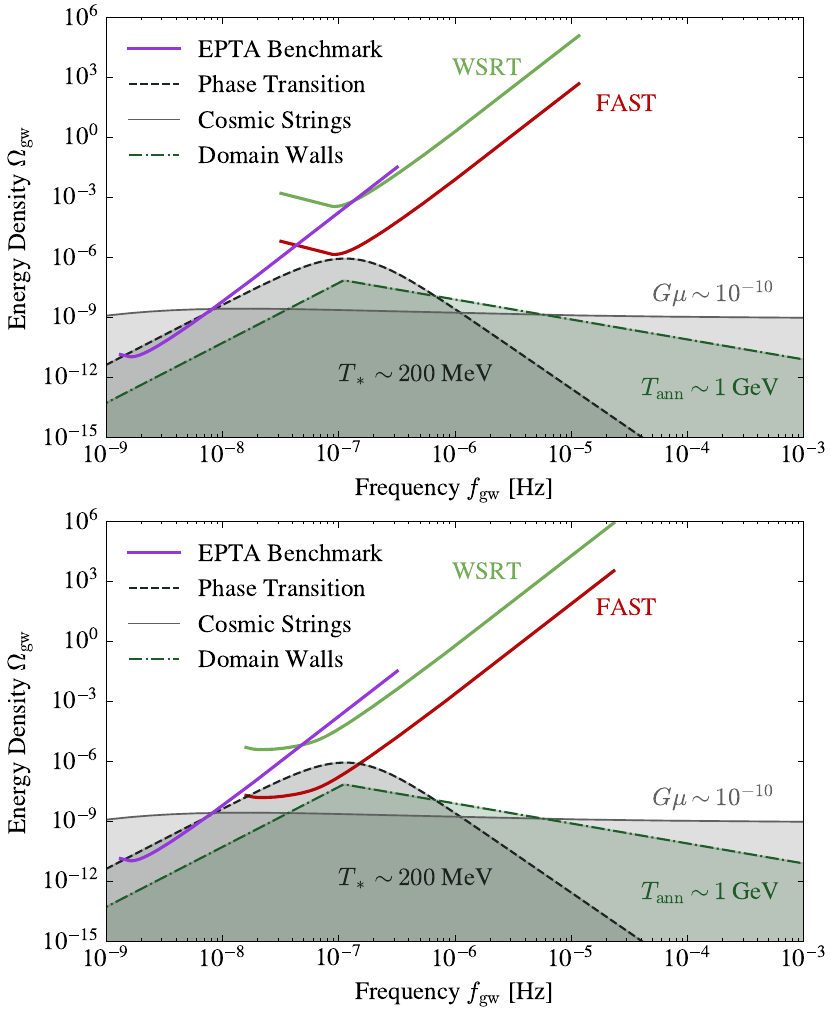}
   \caption{The theoretical minimum energy density sensitivity to a 5-SNR GW background signal of the \textmu Hz-tuned PTA, using both WSRT-like and FAST-like observatories (\textit{top}: Campaign A; \textit{bottom}: Campaign B). This is compared to simulated GWB signals generated by a QCD ($T_*\sim200~\text{MeV}$) phase transition, a cosmic string network with tension $G\mu\sim10^{-10}$ and domain walls with annihilation temperature of $T_\text{ann}\sim1$ GeV. Also included as a benchmark is an idealised sensitivity of EPTA DR2, computed using the same method with the following parameters: 25 pulsars observed with 20~$\text{yr}^{-1}$ cadence over 24 years with a uniform $\sigma_\text{rad}=1~\upmu$s.}
    \label{fig:computed_gw_scurves1}
\end{figure}

In Figure \ref{fig:computed_gw_scurves1} the theoretical sensitivity of the \textmu Hz-tuned PTA, for Campaign A and B, computed using \texttt{hasasia}, is compared to the projected GW energy density of these sources. We also include an EPTA curve as a benchmark for current PTAs (consisting of 25 pulsars observed with a cadence of 20 ${\rm yr}^{-1}$ for 24 years), which is generated using the same radiometer noise-only assumption. We use this simulated data rather than actual data to allow for a fair comparison with the proposed PTA campaigns, considering we omit non-radiometer noise from our PTA models. Considering the Campaign A scenario, the WSRT-like curve reaches a minimum $\Omega_{\rm gw}$ sensitivity of approximately $10^{-4}$, occuring at a GW frequency of $0.1~$\textmu Hz. The FAST-like observatory reaches a sensitivity of $\sim10^{-6}$ at the same frequency. While the WSRT-like curve remains well above the upper limits of all GWB sources, the FAST-like curve gets within an order of magnitude of the QCD phase transition signature. Considering the Campaign B scenario, we see a two-order-of-magnitude improvement to both WSRT and FAST curves. While this has no effect on a WSRT-PTA's ability to constrain the presented cosmological backgrounds, it now performs better than the EPTA benchmark above 0.1 \textmu Hz. The FAST-PTA curve is now almost 4 orders of magnitude better than the EPTA benchmark, and possessing the possibility of constraining both phase transition and domain wall signal. The frequency with best sensitivity has shifted lower, as expected from the larger observation campaign length. While it might be obvious to consider increasing $T_{\rm obs}$ as a means of improving sensitivity, increasing the total observation time to decadal lengths will lower the maximum detectable frequency to nanohertz values in accordance with Equation \eqref{eq:fmax}, which is counter to the goal of probing the microhertz regime. Finally we consider power-law sensitivity of the DSA+SKA-Mid proposal in the left-hand plot in Fig. \ref{fig:dsa-ska}. Comparing it to FAST Campaign B we see that it has the same order of magnitude sensitivity (with $\Omega^{\rm min}_{\rm gw}\sim10^{-8}$), making it capable of probing both the phase transition and domain wall background as modelled in this paper. We expect the DSA+SKA-Mid PTA would perform better than FAST Campaign B had we included non-radiometer noise, as the factor of six more pulsars would allow for better noise modelling.

\subsection{Individual Source Sensitivity}
\begin{figure*}
    \centering
    \includegraphics[width=0.95\linewidth]{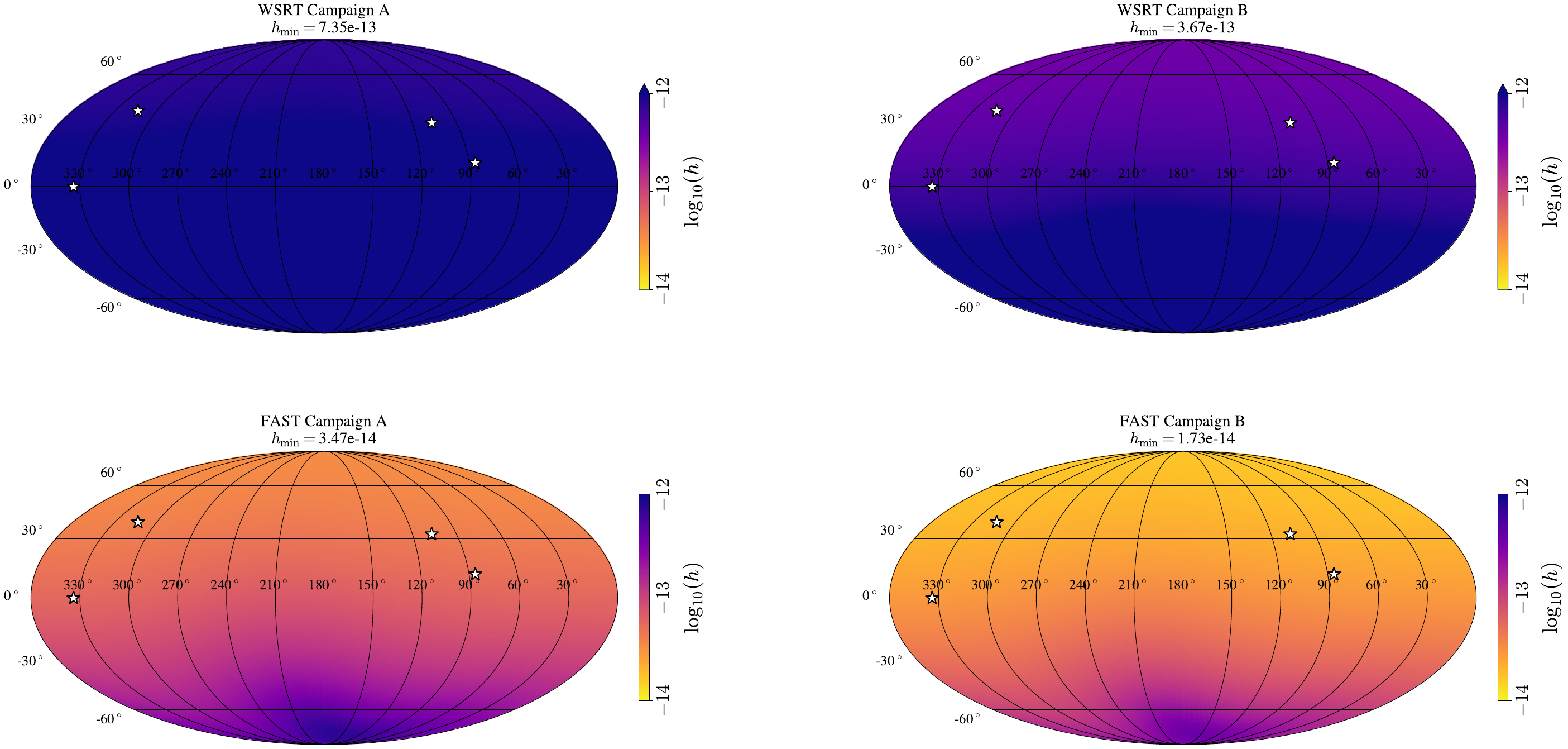}
    \caption{\textit{Top}: Sky map of the ecliptic sphere from the hypothetical four-pulsar timing array using a WSRT-like telescope for the two different campaigns; Campaign A ($T_\text{obs}=1$ yr, 2 ToAs/day) on the left and Campaign B ($T_\text{obs}=2$ yr, 4 ToAs/day) on the right. \textit{Bottom}: The same as above but using a FAST-like telescope instead. All four sky maps display the faintest detectable strain amplitude $h$ of a 5-SNR signal at a GW frequency of $1~$\textmu Hz, for a given campaign. The optimal-direction value, $h_\text{min}$, is also presented. The four pulsars are denoted by white stars.}
    \label{fig:computed_gw_skymap1}
\end{figure*}

We take the strain amplitude upper-limit from individual SMBHBs of $h=8.9\times10^{-14}$ at 1~\textmu Hz from \cite{Zou_2026} as our benchmark for a continuous GW signal in the microhertz regime. Using the effective noise power spectral density computed via \texttt{hasasia} and Equation \eqref{eq:skymap_snr}, we compute the minimum GW strain amplitude $h$ that would have a signal-to-noise ratio of 5, using campaigns A and B on both the WSRT- and FAST-like telescopes (at 1 \textmu Hz). We compute $h$ for each sky position, which is shown in Fig. \ref{fig:computed_gw_skymap1}. The optimal sky-location is predicted, for this set of pulsars, to be latitude +83$^\circ$, longitude 207$^\circ$ on the ecliptic (RA $17^{\rm h}3^{\rm m}$, Dec $+63^\circ$); but as is clear from the colour gradient in Fig. \ref{fig:computed_gw_skymap1}, the whole region 30$^\circ$ above the ecliptic plane shares the same order-of-magnitude sensitivity. Neither campaigns using a WSRT-like telescope would see a GW signal with the strain amplitude, given, as both predict minimum strain of $>3\times10^{-13}$. However a better result is seen in both FAST-like campaigns, with a predicted optimal-direction minimum strain amplitude of $3.47\times10^{-14}$ for Campaign A, and $1.73\times10^{-14}$ for Campaign B. Thus if there was such a target source in the optimal sky location, and it was emitting at the upper-limit strain amplitude established in \cite{Zou_2026}, both FAST campaigns would detect it in this scenario. Admittedly this is a compounding list of best-case scenarios, but as the sky-maps show the $h\sim10^{-14}$ region actually covers a substantial fraction of the sky. At the very least it would likely establish new upper-limits on individual GW sources in the microhertz regime.

\begin{figure*}
    \begin{subfigure}{0.45\linewidth}
        \centering
        \includegraphics[trim={0 0 20cm 0cm},clip,height=5cm]{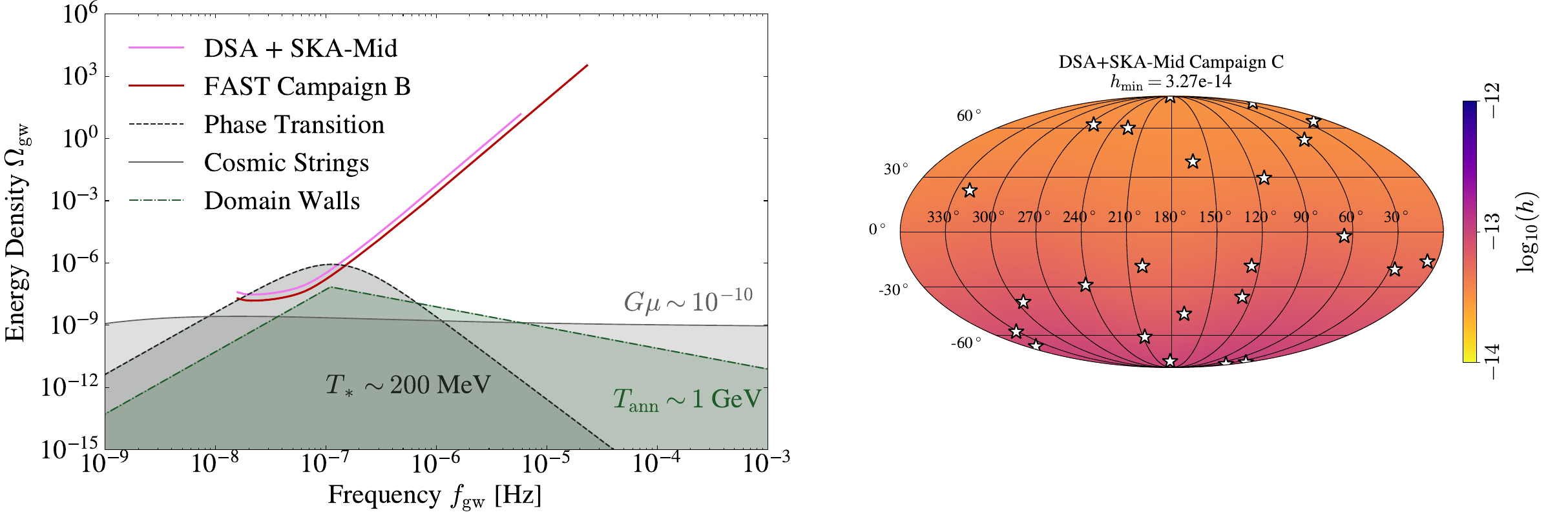}
    \end{subfigure}
    \begin{subfigure}{.45\linewidth}
        \centering
        \includegraphics[trim={21cm 2.5cm 0 1.45cm},clip,height=5cm]{plots/dsa-ska_plcurve-skymap.pdf}
    \end{subfigure}
. 
    \caption{\textit{Left}: The power-law sensitivity of the proposed DSA+SKA-Mid PTA (24 MSPs, 1 ToA per day), compared to FAST Campaign B. Also included are a selection of potential cosmological backgrounds. \textit{Right}: An ecliptic sky-map given in the same units as Fig. \ref{fig:computed_gw_skymap1} using the proposed DSA+SKA-Mid PTA. The pulsars are randomly distributed across the sky.}
    \label{fig:dsa-ska}
    \vspace{0.75cm}
\end{figure*}

The right-hand plot of Fig. \ref{fig:dsa-ska} shows the minimum strain amplitude across sky of a 5-SNR signal detectable by the DSA+SKA-Mid campaign. As the plotted MSPs are just randomly chosen points on the sky, we will ignore the specific optimal-direction, as this will change with an actual choice of pulsars. The best strain amplitude detectable in this idealised scenario is $3.27\times10^{-14}$, which is on par with FAST Campaign A and a factor of about 3 times worse of FAST Campaign B. However like mentioned in \S\ref{sec:rslts_subsec1}, this would likely be different once other noise components are considered. A major advantage of this over just using the singular telescope is the uniform sky-coverage that can be achieved using the Northern-hemisphere DSA and the Southern-hemisphere SKA-Mid.

\section{Conclusion} \label{sec:cnclsion}
As calculated in \S\ref{sec:rslts-discssn}, a 100-m class (WSRT-like) PTA would struggle to provide novel insight into the microhertz GW regime. The power-law sensitivity of Campaign A performs on par with the idealised EPTA at all frequencies $>0.1$ \textmu Hz, and Campaign B is only an order of magnitude better, but remains well above any of the proposed GWB upper-limits. It also would not be able to detect an individual source emitting GWs at the upper-limit established in \cite{Zou_2026}. The 500-m class (FAST-like) PTA campaigns are both much more promising, with both campaigns having a better power-law sensitivity than the white-noise-only EPTA curve in the $>0.1~$\textmu Hz range. Thus it is likely a FAST-like Campaign B would be able to set new upper limits on the cosmological backgrounds mentioned. Likewise it could detect an individual source at the upper-limit of $8.9\times10^{-14}$ for Campaigns A and B respectively, and even probe deeper, meaning it could also establish new upper bounds for individual GW sources in the ~\textmu Hz regime. An important factor to consider however is the telescope time such a campaign would require. Over the course of 2 years, campaign B would require 11,680 hours of telescope time. Given the competition for telescope time on an instrument such as FAST, such an observational campaign is incredibly unlikely. Comparatively, a DSA and SKA-Mid \textmu Hz PTA would require approximately 1,095 hours for each telescope over the same two-year timespan, and would achieve similar results to the FAST Campaign B. Furthermore, the DSA has already committed to this order of magnitude pulsar timing cadence \citep{Hallinan_2019}, and we believe SKA-Mid capable of it likewise; thus it is a matter of when, not if, timing data like these become available. It is important to reiterate that this is an idealised-noise study, and as such the results must be understood as representing the most optimistic performance of these proposed PTAs. Further work must be done to investigate the noise environment of the microhertz regime and its effect on the sensitivities of each campaign discussed above, in preparation for timing data from telescopes like the DSA and SKA. 

Any ToAs generated in a microhertz-targeting campaign are reusable, as they can be combined with other datasets, like the International Pulsar Timing Array, meaning timing experiments like this have use outside of microhertz-specific searches. High-cadence subsets are of value to the PTA community as they have already been used to put novel constraints on individual sources of GWs \citeg{Caballero_2025,Zou_2026}. Despite these efforts, the microhertz regime still stands as one of the least-explored regions of the gravitational wave spectrum. There are clear science motivators to explore this regime, including the aforementioned cosmological sources and bright individual supermassive black hole binaries. Thus as we await the development of new GW detectors targeting microhertz GW frequencies \citeg{muAres_2021,Fedderke_2022}, the boundaries of pulsar timing should be pushed and tested to its limit.

\section*{Acknowledgements}
We thank Gemma Janssen for supplying information on WSRT. We also thank Boris Goncharov for taking the time to review this manuscript and for providing valuable feedback. This work has emanated from research conducted with the financial support of Taighde Éireann - Research Ireland under Grant
number GOIPG/2025/5806.

\section*{Code \& Data Availability}
The power-law sensitivity curves, sky sensitivity maps and all other related data, excluding the cosmological background data, are deposited at \href{https://doi.org/10.5281/zenodo.23690555}{\color{cyan}{10.5281/zenodo.23690555}} \citep{dataset}. All code written for this work is available at the GitHub repository \href{https://github.com/consher/microhertz_PTA}{\color{cyan}{consher/microhertz\_PTA}}.

This research made use of the following libraries: \texttt{NumPy} \citep{numpy}, \texttt{pandas} \citep{pandas}, \texttt{Matplotlib} \citep{matplotlib}, \texttt{HealPy} \citep{healpy1,healpy2}, \texttt{Hasasia} \citep{Hazboun_2019a}, \texttt{SciencePlots} \citep{SciencePlots}, \texttt{AstroPy} \citep{astropy}, \texttt{BeautifulSoup} and \texttt{Requests}.

\bibliographystyle{apsrev4-1}

\bibliographystyle{aasjournal}
\bibliography{references}

\end{document}